\newif\ifUsenix\Usenixfalse 
\newif\ifAnon\Anonfalse 

\newcommand{\Func}[1]{\texttt{#1}}
\newcommand{\Path}[1]{\textit{#1}}

\ifUsenix
  \documentclass[letterpaper,twocolumn,10pt]{article}
  \usepackage{usenix2019_v3}
  \usepackage{authblk}
\else
  \documentclass[conference]{IEEEtran}
\fi
\usepackage{tugraz_defaults}
\usepackage{subfigure}

\usepackage{amsmath}
\usepackage{amssymb}
\usepackage{mathtools}
\usepackage{multirow}
\newcolumntype{P}[1]{>{\centering\arraybackslash}p{#1}}
\newcolumntype{M}[1]{>{\centering\arraybackslash}m{#1}}
\usepackage{tablefootnote}

\makeatletter
\newcommand\footnoteref[1]{\protected@xdef\@thefnmark{\ref{#1}}\@footnotemark}
\makeatother

\usepackage{hyperref}
    \definecolor{linkcolor}{rgb}{0,0.1,0.35}
    \definecolor{citecolor}{rgb}{0,0.4,0}
    \definecolor{urlcolor}{rgb}{0,0,0.65}
    \hypersetup{pdfpagemode=UseNone,pdfstartview=FitH,colorlinks=true, linkcolor=linkcolor, urlcolor=urlcolor, citecolor=citecolor}

\definecolor{dkgreen}{rgb}{0,0.6,0}
\definecolor{gray}{rgb}{0.5,0.5,0.5}
\definecolor{mauve}{rgb}{0.58,0,0.82}
\usepackage[toc,acronym,nowarn]{glossaries}

\newacronym{OS}{OS}{Operating System}
\newacronym{SE}{SE}{Secure Element}
\newacronym{NFC}{NFC}{Near Field Communication}
\newacronym{BLE}{BLE}{Bluetooth Low Energy}
\newacronym{RIL}{RIL}{Radio Interface Layer}
\newacronym{SELinux}{SELinux}{Security-Enhanced Linux}
\newacronym{LSM}{LSM}{Linux Security Modules}
\newacronym{TCB}{TCB}{Trusted Computing Base}
\newacronym{MAC}{MAC}{Message Authentication Code}
\newacronym{IPC}{IPC}{Inter-Process Communication}
\newacronym{ICC}{ICC}{Inter-Container Communication}
\newacronym{cgroups}{cgroups}{control groups}
\newacronym{CA}{CA}{Certificate Authority}
\newacronym{PKI}{PKI}{Public Key Infrastructure}
\newacronym{MITM}{MITM}{Man-In-The-Middle}
\newacronym{BYOD}{BYOD}{Bring-Your-Own-Device}
\newacronym{CM}{CM}{Container Management}
\newacronym{SM}{SM}{Security Management}
\newacronym{HAL}{HAL}{Hardware Abstraction Layer}
\newacronym{TLS}{TLS}{Transport Layer Security}
\newacronym{C2C}{C2C}{Container To Container}
\newacronym{Protobuf}{Protobuf}{Protocol Buffers}
\newacronym{SDO}{SDO}{Sensitive Data Object}
\newacronym{VMA}{VMA}{Virtual Memory Area}
\newacronym{PGD}{PGD}{Page Global Directory}
\newacronym{PUD}{PUD}{Page Upper Directory}
\newacronym{PMD}{PMD}{Page Middle Directory}
\newacronym[firstplural=Page Table Entries (PTEs)]{PTE}{PTE}{Page Table Entry}
\newacronym{COW}{COW}{Copy-On-Write}
\newacronym{IV}{IV}{Initialization Vector}
\newacronym{ESSIV}{ESSIV}{Encrypted Salt-Sector Initialization Vector}
\newacronym{KSM}{KSM}{Kernel Samepage Merging}
\newacronym{JIT}{JIT}{Just-In-Time}
\newacronym{DMA}{DMA}{Direct Memory Access}
\newacronym{FDE}{FDE}{Full Disk Encryption}
\newacronym{AS}{AS}{Address Space}
\newacronym{GCM}{GCM}{Google Cloud Messaging}
\newacronym{TPM}{TPM}{Trusted Platform Module}
\newacronym{JTAG}{JTAG}{Joint Test Action Group}
\newacronym{LUKS}{LUKS}{Linux Unified Key Setup}
\newacronym{VPN}{VPN}{Virtual Private Network}
\newacronym{PBKDF2}{PBKDF2}{Password-Based Key Derivation Function 2}
\newacronym{KVM}{KVM}{Kernel-based Virtual Machine}
\newacronym{VM}{VM}{Virtual Machine}
\newacronym{firstplural=Virtual Machines}{VM}{VM}
\newacronym{HV}{HV}{Hypervisor}
\newacronym{SEV}{SEV}{Secure Encrypted Virtualization}
\newacronym{SME}{SME}{Secure Memory Encryption}
\newacronym{RMP}{RMP}{Reverse Map Table}
\newacronym{TSME}{TSME}{Transparent \gls{SME}}
\newacronym{SP}{SP}{Secure Processor}
\newacronym[firstplural=Guest Virtual Addresses (GVAs)]{GVA}{GVA}{Guest Virtual Address}
\newacronym[firstplural=Guest Physical Addresses (GPAs)]{GPA}{GPA}{Guest Physical Address}
\newacronym[firstplural=Host Physical Addresses (HPAs)]{HPA}{HPA}{Host Physical Address}
\newacronym{GPT}{GPT}{Guest Page Table}
\newacronym{HPT}{HPT}{Host Page Table}
\newacronym{TLB}{TLB}{Translation Lookaside Buffer}
\newacronym{PoC}{PoC}{Proof of Concept}
\newacronym{ORAM}{ORAM}{Oblivious RAM}
\newacronym{SEV-ES}{SEV-ES}{SEV Encrypted State}
\newacronym{VMCB}{VMCB}{Virtual Machine Control Block}
\newacronym{SLAT}{SLAT}{Second Level Address Translation}
\newacronym[firstplural=Nested Page Tables (NPTs)]{NPT}{NPT}{Nested Page Table}
\newacronym{SSH}{SSH}{Secure Shell}
\newacronym{RSA}{RSA}{Rivest–Shamir–Adleman}
\newacronym{ECDHE}{ECDHE}{Elliptic-Curve Diffie-Hellman Ephemeral}
\newacronym{AES}{AES}{Advanced Encryption Standard}
\newacronym{ECB}{ECB}{Electronic Codebook Mode}
\newacronym{OOM}{OOM}{Out Of Memory}
\newacronym{TME}{TME}{Total Memory Encryption}
\newacronym{MKTME}{MKTME}{Multi-Key Total Memory Encryption}
\newacronym{VMI}{VMI}{Virtual Machine Introspection}
\newacronym{MAD}{MAD}{Median Absolute Deviation}
\newacronym{HSM}{HSM}{Hardware Security Module}
\newacronym{AE}{AE}{Automatic Exit}
\newacronym{NAE}{NAE}{Non-Automatic Exit}
\newacronym{AES-NI}{AES-NI}{AES New Instructions}
\newacronym{NIC}{NIC}{Network Interface Card}
\newacronym{NMI}{NMI}{Non-Maskable Interrupt}
\newacronym{MTU}{MTU}{Maximum Transmission Unit}
\newacronym[firstplural=Virtual Addresses]{VA}{VA}{Virtual Address}
\newacronym[firstplural=Physical Addresses]{PA}{PA}{Physical Address}
\newacronym{GFN}{GFN}{Guest Frame Number}
\newacronym{IOMMU}{IOMMU}{I/O Memory Management Unit}
\newacronym{AISE}{AISE}{Address Independent Seed Encryption}
\newacronym{MT}{MT}{Merkle Tree}
\newacronym{BMT}{BMT}{Bonsai Merkle Tree}
\newacronym{LPID}{LPID}{Located Page IDentifier}
\newacronym{CB}{CB}{Counter Block}
\newacronym{PRD}{PRD}{Page Root Directory}
\newacronym{SWIOTLB}{SWIOTLB}{Software I/O Translation Buffer}
\newacronym{ASID}{ASID}{Address Space Identifier}
\newacronym{vCPU}{vCPU}{virtual CPU}
\newacronym{GHCB}{GHCB}{Guest Hypervisor Communication Block}
\newacronym{XEX}{XEX}{Xor-Encrypt-Xor}
\newacronym{XE}{XE}{Xor-Encrypt}
\newacronym{SGX}{SGX}{Software Guard Extensions}
\newacronym{KASLR}{KASLR}{Kernel Address Space Layout Randomization}
\newacronym{VC}{\texttt{\#VC}}{VMM Communication Exception}
\newacronym{IBS}{IBS}{Instruction Based Sampling}
\newacronym{SEV-SNP}{SEV-SNP}{SEV Secure Nested Paging}
\newacronym{CCA}{CCA}{Chosen Ciphertext Attack}
\newacronym{CPA}{CPA}{Chosen Plaintext Attack}

\makeglossaries
\glsdisablehyper

\begin{document}

\title{
	\Large \bf SEVurity: No Security Without Integrity\\\large\bf Breaking Integrity-Free Memory Encryption with Minimal Assumptions
}

\ifAnon  
  \author{Anonymous Submission}
\else
  \ifUsenix
    \author[1]{Luca Wilke}
    \author[1]{Jan Wichelmann}
    \author[2]{Mathias Morbitzer}
    \author[1]{Thomas Eisenbarth}
    \affil[1]{University of L\"ubeck, L\"ubeck, Germany}
    \affil[2]{Fraunhofer AISEC, Garching, Germany}
  \else
    \author{
      \IEEEauthorblockN{
        Luca Wilke\IEEEauthorrefmark{1},
        Jan Wichelmann\IEEEauthorrefmark{1},
        Mathias Morbitzer\IEEEauthorrefmark{2},
        Thomas Eisenbarth\IEEEauthorrefmark{1}
    }
    \IEEEauthorblockA{\IEEEauthorrefmark{1}University of L\"ubeck, L\"ubeck, Germany\\
    \{l.wilke,j.wichelmann,thomas.eisenbarth\}@uni-luebeck.de}
    \IEEEauthorblockA{\IEEEauthorrefmark{2}Fraunhofer AISEC, Garching, Germany\\
    \{mathias.morbitzer\}@aisec.fraunhofer.de}
  }
  \fi
\fi

\maketitle
\thispagestyle{plain}
\pagestyle{plain}

\begin{abstract}
    One reason for not adopting cloud services is the required trust in the cloud provider: As they control the hypervisor, any data processed in the system is accessible to them. Full memory encryption for Virtual Machines (VM) protects against curious cloud providers as well as otherwise compromised hypervisors. AMD Secure Encrypted Virtualization (SEV) is the most prevalent hardware-based full memory encryption for VMs. Its newest extension, SEV-ES, also protects the entire VM state during context switches, aiming to ensure that the host neither learns anything about the data that is processed inside the VM, nor is able to modify its execution state. Several previous works have analyzed the security of SEV and have shown that, by controlling I/O, it is possible to exfiltrate data or even gain control over the VM's execution.
    In this work, we introduce two new methods that allow us to inject arbitrary code into SEV-ES secured virtual machines. Due to the lack of proper integrity protection, it is sufficient to reuse existing ciphertext to build a high-speed encryption oracle. As a result, our attack no longer depends on control over the I/O, which is needed by prior attacks. As I/O manipulation is highly detectable, our attacks are stealthier. In addition, we reverse-engineer the previously unknown, improved Xor-Encrypt-Xor (XEX) based encryption mode, that AMD is using on updated processors, and show, for the first time, how it can be overcome by our new attacks.

\end{abstract}

\section{Introduction}
\glspl{VM} are a very important part of today's cloud computing market.
They significantly ease technical aspects of hosting. On the customer side they allow for flexible resource scaling, due to their low setup time. Furthermore, multiple \glspl{VM} can run on the same physical machine, because the hypervisor -- the software that manages the virtualization -- provides one sided isolation, by preventing the \gls{VM} from accessing other software running on the host. 
Thus, the hosting provider can make better use of its hardware.

Apart from that, potential customers still cite data privacy concerns toward cloud service providers as a main reason not to adopt cloud solutions, especially in cases where the hosting location within a given jurisdiction cannot be guaranteed. Performing sensitive computations in \glspl{VM} requires the customer to fully trust the hypervisor, since the hypervisor has direct access to all virtualized resources.

Security solutions like full disk encryption only partially address this issue, since the data is still vulnerable when being decrypted and stored in the RAM at run time.

Providing full isolation between the hypervisor and the \gls{VM} has been 
studied extensively by researchers as well as industry \cite{anati2013innovative,kaplan2016amd,sev-es,jin2011architectural,xia2013architecture,intelTME}.
Intel \gls{SGX} \cite{anati2013innovative,hoekstra2013using,mckeen2013innovative} was the first widely available solution for protecting data in RAM. 
However, it only can protect a small chunk of RAM, not the \gls{VM} as a
whole~\cite{gueron2016memory}.
In 2016, AMD introduced \gls{SME} and \gls{SEV}~\cite{kaplan2016amd} to protect the entire system memory. \gls{SME} provides drop-in,
AES-based RAM encryption.
\gls{SEV} extends this for \glspl{VM} by using different encryption keys per \gls{VM}, in order to prohibit the hypervisor from inspecting the \gls{VM}'s main memory.
This was a first step towards full isolation. The Linux kernel support for \gls{SEV} was mainlined in early 2018~\cite{amdWebsite}. In February 2017, AMD introduced \gls{SEV-ES} \cite{sev-es}, which offers additional protection against manipulating the state of a \gls{VM} during
context switches. While \gls{SEV-ES} does not need new hardware, it requires extensive modifications to the Linux kernel. According to AMD, the corresponding patches are mostly finished, however 
support for \gls{SEV-ES} has not been mainlined, yet~\cite{lendacky2019}.
Intel is also working on a solution similar to \gls{SME}/\gls{SEV}, 
called \gls{TME}/\gls{MKTME}~\cite{intelTME}, but did not yet publish corresponding processors.
A detailed comparison between Intel \gls{SGX} and AMD's memory encryption
can be found in~\cite{mofrad2018comparison}.

This work focuses on AMD's solutions for providing full isolation between hypervisor and \gls{VM}, as it 
is the most prevalent full memory encryption.
All prior attacks have either been mitigated by \gls{SEV-ES} or used I/O to move known plaintext into encrypted 
pages. 
We show that our attack vector is available even without user-controlled I/O or access to unprotected I/O operations.
Instead, we only require minimal knowledge about the system to compromise and completely take
over the \gls{VM}.
To achieve this, we bootstrap an encryption oracle from just a few megabytes of known plaintext, allowing us to place and execute arbitrary code in the \gls{VM}.
We identify the lack of integrity protection as the main reason for this weakness, and postulate that full security against our attacks can only be achieved by implementing a proper integrity protection scheme.
As an independent contribution, we also show that AMD's updated XEX-based memory encryption mode is still vulnerable to the previous attacks.

\subsection{Our Contribution}
\begin{itemize}
    \item We exploit the missing integrity protection of SEV to place arbitrary code in a \gls{SEV-ES} secured \gls{VM}, \emph{without relying on any I/O operations}.
    
    \item For this we bootstrap an encryption oracle just by moving existing ciphertext within the VM's memory.
    
    \item We show the security impact of the emulated \texttt{cpuid} instruction by abusing it to create a high-performance encryption oracle.
    
    \item We reverse engineer the new XEX encryption mode, that is used on updated processors, and infer the associated tweak values. We show that this mode is just as vulnerable as the previous XE-based encryption mode which has been exploited by prior attacks.
    
    \item We discuss previously proposed and new countermeasures, and evaluate their impact on our attacks.
\end{itemize}

\section{Background} \label{sec:background}

\subsection{AMD Memory Encryption} \label{ssec:amd-memory-encryption}

\noindent\textbf{Secure Memory Encryption (SME)}
To protect against an attacker with physical access to a system, AMD introduced \gls{SME} in 2016 \cite{kaplan2016amd}. 
\gls{SME} encrypts data before writing it to RAM, which renders it useless for an attacker attempting to access the data, e.g., via a cold boot attack or \gls{DMA} \cite{halderman2008lest,bauer2016lest,yitbarek2017cold}.
The encryption and decryption are controlled by the \gls{SP}, an ARM-based co-processor. 

A special bit in the page table -- the so-called C-bit -- is used to indicate whether a page should be encrypted \cite{kaplan2016amd}. However, changing the C-bit does not change the content of the page, only whether it is
interpreted as encrypted or not. There is no coherency between mappings of the same memory location
with different C-bit values, or different encryption keys. Thus, changing the encryption status requires flushing all involved CPU caches.
In case the \gls{OS} does not support \gls{SME}, \gls{TSME} can be used: In \gls{TSME} mode, all memory pages are encrypted independently from the value of the C-Bit. 

When the system boots, a random memory encryption key is created and stored in the \gls{SP}~\cite{kaplan2016amd}. Subsequently, memory writes are encrypted, and memory reads are decrypted. For both the encryption and decryption, the \gls{SP} uses AES in conjunction with a physical address-based tweak.

\smallskip\noindent\textbf{Secure Encrypted Virtualization (SEV)} The \gls{SEV} technology was introduced together with \gls{SME} in 2016.
The goal of \gls{SEV} is to protect a \gls{VM} from a malicious or compromised hypervisor.
This is achieved by using the memory encryption technology of \gls{SME} with a different encryption key for each \gls{VM} and the hypervisor itself. All keys are stored in the \gls{SP} and are not accessible by any other party. 

Due to the different encryption keys, a hypervisor attempting to access data of a \gls{VM} would get a decrypted version using the hypervisor's key, rendering the generated plaintext useless. However, the \glspl{VM} can use the hypervisor's key to intentionally share information.

A known issue of \gls{SEV} is the lack of encrypting the \gls{VMCB}, which is a data structure describing the state of a \gls{VM}.
It includes information like the VM's configuration and register contents. It also provides means for communication between the hypervisor and the \gls{VM}: For example, if the \gls{VM} exits due to an interrupt, the processor stores appropriate metadata (e.g., a memory address) in this structure. The lack of encryption can be exploited to manipulate the execution flow of the \gls{VM} and leak sensitive information like \cite{hetzelt2017security, werner2019severest} have shown.

\smallskip\noindent\textbf{SEV Encrypted State (SEV-ES)} To address these issues, AMD introduced \gls{SEV-ES}~\cite{sev-es} as an extension for \gls{SEV}.
\gls{SEV-ES} splits the \gls{VMCB} into two areas: The control area and the save area. The unencrypted control area contains information that must always be available to the hypervisor in order to manage the \gls{VM}, e.g., flags for interrupt injection. The save area contains all of the other information from the \gls{VMCB}, and is protected against access or manipulation from the hypervisor by encrypting it when the \gls{VM} exits. However, since certain operations require the \gls{VM} to share data from its save area with the hypervisor (e.g., reading and writing certain registers when emulating \texttt{cpuid}), AMD introduced the \gls{GHCB}, which basically is a shared page, allowing communication between guest and hypervisor. They introduced a new exception, which gets triggered by operations that require the \gls{VM} to share information with the hypervisor, allowing the guest to copy the required data from the \gls{VMCB} to the \gls{GHCB} before the \texttt{\#VMEXIT}. When the \gls{VM} is resumed, it can copy the data back to its \gls{VMCB}.

\smallskip\noindent\textbf{Encryption mode}
Like \gls{SME}, \gls{SEV} and \gls{SEV-ES} provide drop-in memory encryption, but for \glspl{VM}. The memory encryption has no ciphertext expansion, which means that structure and size of the memory remain unchanged with and without encryption. Similar techniques like Intel SGX~\cite{costan2016intel} store \gls{MAC} tags for each memory block, which allows for strong integrity protection, but comes with significant overhead. In contrast, AMD does not store any integrity protecting metadata, which is very convenient for the user in terms of transparency and space-efficiency.

AMD achieves an implicit block level integrity protection through the encryption: Changing any bits in a ciphertext block results in a garbled and for the attacker unpredictable plaintext block. Also, the usage of an address-based tweak should make it difficult to decrypt a valid ciphertext at another address and get a meaningful plaintext. One thus has to assume that the VM execution will eventually halt if it encounters random data blocks (i.e. invalid opcodes or state variables) -- there are no means of reliably detecting whether the ciphertext has been tampered with.

Since both \gls{SME} and \gls{SEV} use the same technique for the encryption process, they suffer from the same problems regarding integrity. These problems are even more severe in the case of \gls{SEV}, because an attacker with hypervisor permissions can easily manipulate or copy the RAM content of a \gls{VM}.

\subsection{Memory Encryption using Tweakable Block Ciphers}
One popular method for storage encryption are tweakable block ciphers, such as AES-XTS~\cite{XTSAES}, which is, e.g., used in Apple's FileVault, MS Bitlocker and Android's file-based encryption. Tweakable block ciphers provide encryption without data expansion as well as some protection against plaintext manipulation. A \emph{tweak} allows to securely change the behavior of the block cipher, similar to instantiating it with a new key, but with little overhead. If the tweak is, e.g., a function of the block address, the encryptions for each block appear independent, which prevents numerous attacks on the ciphertext, such as frequency analysis, block moving and several more. However, without proper integrity protection, several attacks remain possible, i.e., randomizing the plaintext (by altering the ciphertext), replaying old values and traffic analysis by monitoring location-specific changes.

AES-XTS includes ciphertext stealing, which allows expansion-free block encryption for arbitrary-length plaintexts by using previous ciphertext for padding.
Memory in RAM and the uncore part of the CPU is always handled in blocks of 64 bytes, which is a multiple of the 16 byte block size of AES. Thus,
ciphertext stealing is not needed, reducing the XTS mode to the original \gls{XEX} mode by Rogaway~\cite{rogaway2004efficient}.
\gls{XEX} and \gls{XE} are methods to turn a block cipher such as AES into a tweakable blockcipher, where a tweak-derived value is XORed with the plaintext before encryption (and XORed again after encryption in the case of \gls{XEX}).

\subsection{Nested Paging}
On most common \glspl{OS}, processes use \glspl{VA} to access data~\cite{Drepper2007virtual}. 
Those \glspl{VA} are translated into \glspl{PA}, which determine where the data is located in the physical memory. 
The mappings between \glspl{VA} and \glspl{PA} are stored in the page table.

On virtualized systems, two different page tables are used. 
Within the \gls{VM}, the \gls{VA} used by the guest, the \gls{GVA}, is translated to the \gls{GPA}. 
The \gls{GPA} is the address which the \gls{VM} considers to be the \gls{PA}. 
However, on the host system itself another page table is introduced, to allow multiple \glspl{VM} to run on the same physical machine. 
This second page table is called the \gls{SLAT}, or \gls{NPT}~\cite{AMD2008nested}. 
The \gls{NPT} translates the \gls{GPA} into the \gls{HPA}, the actual address of the data in physical memory. 

When \gls{SEV} is active, the page table in the guest is encrypted and thus not accessible by the hypervisor. 
However, the hypervisor is still responsible for managing the \gls{NPT}. 
This allows the hypervisor to infer information about the \gls{VM}'s memory assignment by monitoring the entries in the \gls{NPT}. 
The \gls{NPT} cannot be accessed by the \gls{VM}. 
It is therefore not possible for the \gls{VM} to prevent the hypervisor from overwriting permissions in the \gls{NPT}. 
Multiple attacks make use of this possibility to gather information about where the \gls{VM} stores critical data \cite{hetzelt2017security, morbitzer2018severed, buhren2018detectability, li2019exploiting, morbitzer2019extracting}.

\subsection{Instruction Interception}\label{sec:interception}

In a virtualized environment, there are two reasons which cause a \gls{VM} to trigger a \texttt{\#VMEXIT}, which hands
control back to the hypervisor. 
One reason are interrupts and exception handlers, which need hypervisor assistance, e.g., page faults due to swapped pages.

The second reason is the interception of special instructions \cite{AMD2019architecture}. Two prominent examples for this are the \texttt{cpuid} and the \texttt{rdtsc} instruction.
The \texttt{cpuid} instruction allows querying a wide span of CPU information, including an accurate model number, a list of supported features and the system's topology. Changing the returned registers allows the hypervisor fine grained control over the hardware features it exposes to the guest.
The \texttt{rdtsc} instruction returns the current state of the core-private timestamp counter. \glspl{OS} and other programs may use this counter for cycle-level time measurements. If a \gls{VM} is live-migrated from one host to another, the \texttt{cpuid} and \texttt{rdtsc} values on both machines might be different. Emulating these instructions allows the hypervisor to convey a consistent picture of the system state.
Whether an instruction is intercepted or not can be configured in the \gls{VMCB}.

\subsection{Previous Attacks on SEV}
\smallskip\noindent\textbf{Manipulating VMCB}
Hetzelt and Buhren \cite{hetzelt2017security} explore the idea of manipulating the general
purpose registers stored in the \gls{VMCB} to create an encryption/decryption oracle.
In order to move data from memory into a register or vice versa, they manipulate the 
\texttt{RIP} register in the \gls{VMCB}, to construct corresponding gadgets. \gls{SEV-ES} mitigates these attacks.

\smallskip\noindent\textbf{I/O-based attacks}
Du et al.~\cite{du2017secure} build an encryption oracle, which is based on self-generated network
traffic. They require that an Nginx web server is running in the \gls{VM} and exploit its memory
management behavior, allowing them to locate the content of specifically crafted, self-generated 
HTTP packets in the \gls{VM}'s RAM.

Morbitzer et al.~\cite{morbitzer2018severed} leverage the hypervisor's control over the
\gls{NPT} in order to swap \gls{GPA} mappings. In combination with a network service running inside the \gls{VM}, which
returns some resources on request, they build a decryption oracle.
In the first phase, they locate the \gls{GPA} where the response of the network service is stored by repeatedly
sending requests and monitoring the page fault side channel. In the second phase, they manipulate the
\gls{NPT} so that the \gls{GPA} of the returned resource points to another memory location. 
Thus, the content of this memory location is returned on the next request.
In their follow-up work \cite{morbitzer2019extracting}, they show how to locate \glspl{GPA} that might contain
secret data, like encryption keys.

Li et al.~\cite{li2019exploiting} exploit
the fact that DMA operations issued by the \gls{VM} are currently performed via an unencrypted bounce buffer.
They demonstrate that this can be used in combination with network I/O, to create an encryption/decryption
oracle. If the \gls{VM} performs network I/O, the packets are copied to the bounce buffer, before they are processed by the
network card. To create an encryption oracle, they manipulate incoming data in the bounce buffer before the \gls{VM} copies
it into its private memory. For the decryption oracle, they manipulate the data that the \gls{VM} wants to send before
it gets copied from the \gls{VM}'s private memory into the bounce buffer. To detect the memory locations and
hit the correct timing, they use the page fault side channel.

\smallskip\noindent\textbf{Fingerprinting Applications} Werner et al.~\cite{werner2019severest}
showed two independent results.
First, they use the unencrypted \gls{VMCB} to reconstruct  code executed in the
\gls{VM} by singlestepping the \gls{VM} while observing the changes to the unencrypted register values in the \gls{VMCB}.
Like \cite{hetzelt2017security} they also use the unencrypted \gls{VMCB} to encrypt/decrypt data.
This is mitigated by \gls{SEV-ES}.
In their second result, they show how to use a performance counter subsystem called
\gls{IBS} to detect which applications are running
in the \gls{VM}. They leverage that \gls{IBS} leaks the \gls{GVA}
of return statements, and show, that the distance between return statements uniquely identifies
specific versions of applications.
They claim that the guest cannot detect whether \gls{IBS} is activated. This result holds under \gls{SEV-ES}.

\smallskip\noindent\textbf{Security of AMD-SP}
In \cite{buhren2019insecure} Buhren et al. take another attack vector. 
They examine the security of the AMD \gls{SP}, which forms the root of
trust for \gls{SEV}. The \gls{SP} only executes signed firmware images. However, they found a bug
in the signature check mechanism, allowing them to execute manipulated firmware on the \gls{SP}. While
newer firmware versions fix that bug, there is no rollback prevention mechanism. Thus an attacker can just load a vulnerable firmware version. Using a modified firmware, they are able to extract the private key, used by the \gls{SP} to authenticate itself as an AMD
device.

\smallskip\noindent\textbf{Data Faults}
In \cite{buhren2017fault}, Buhren et al. explore the idea of performing classical fault attacks on application data in memory. They flip a bit in a ciphertext block, in order
to create garbled plaintext. They demonstrate how to use this to perform a fault attack on RSA CRT. 
They implemented their attack for \gls{SME} and required
that the attacker is able to run an unprivileged application and can perform DMA memory access.
However, it should also be possible to migrate this attack to \gls{SEV}.

\section{Reverse Engineering the Encryption Mode} \label{sec:reversing-the-encryption-mode}
In order to predict how the plaintext that corresponds to a ciphertext block changes, when
the ciphertext block gets copied to a new memory location, we reverse engineer the AES encryption mode, particularly the address-based tweak function. Only with this knowledge we are able to inject meaningful data into the
\gls{VM} via ciphertext moving.

As shown in~\cite{du2017secure}, AMD uses a tweaked AES encryption to avoid that a ciphertext block appears multiple times due to an identical plaintext. If AMD would not have added any randomization, it would have been trivial to move ciphertext blocks, and easy to fingerprint applications by detecting certain repeating patterns in memory, e.g., alignment bytes between functions, or zeroed pages.

Since an encrypted block does not have any kind of tag or temporal information, AMD uses a function of its physical address to compute the associated tweak value. In the following, we summarize our findings on that function and verify and extend the results from~\cite{du2017secure}.

\subsection{XE Encryption Mode} 

According to~\cite{du2017secure}, the processor
contains a fixed array of 16-byte \emph{tweak constants} $t_i$ for $i\geq 4$. Given a physical address $p$, where $\mathrm{bit}(p,i)$ represents its $i$-th least significant bit for $i\geq 0$, the \emph{tweak value} $T(p)$ is defined as
\begin{equation*}
    T(p)\coloneqq\bigoplus^{n-1}_{i=4}\mathrm{bit}(p,i)\cdot t_i,
\end{equation*}
This means, that for each physical address bit the respective tweak constant is XORed, if that bit is 1.

A 16-byte plaintext block $m\in\{0,1\}^n$ with physical address $p$ is then encrypted as
\begin{equation*}
    \mathrm{Enc}_K(m,p)\coloneqq \mathrm{AES}_K\left(m\oplus T(p)\right).
\end{equation*}
Similarly, decryption of a ciphertext $c$ uses the inverse transformation
\begin{equation*}
    \mathrm{Dec}_K(c,p)\coloneqq \mathrm{AES}^{-1}_K(c)\oplus T(p).
\end{equation*}
This construction is a variant of the \gls{XE} mode of operation~\cite{rogaway2004efficient}.

We can exploit the missing integrity protection, to compute all tweak constants $t_i$: We encrypt a block $m$ with physical address $p$, copy the ciphertext to other addresses $q_j$ and decrypt it there. By doing this, the tweak values of the source address $p$ and the target addresses $q_j$ are XORed:
\begin{align*}
    \mathrm{Dec}_K&\left(\mathrm{Enc}_K\left(m,p\right),q_j\right)\\
    &=\mathrm{AES}^{-1}_K\left(\mathrm{AES}_K\left(m\oplus T(p)\right)\right)\oplus T(q_j)\\
    &=m\oplus T(p)\oplus T(q_j)\\
    &=m\oplus T(p\oplus q_j) . 
\end{align*}
This allows us to build a system of linear equations, whose solution are the tweak constants:
\begin{equation*}
    \begin{pmatrix}
        p\oplus q_1\\
        p\oplus q_2\\
        \vdots \\
        p\oplus q_{n-4}
    \end{pmatrix}
    \cdot
    \begin{pmatrix}
        t_{n-1}\\
        t_{n-2}\\
        \vdots\\
        t_{4}
    \end{pmatrix}
    =
    \begin{pmatrix}
        m\oplus T(p\oplus q_1)\\
        m\oplus T(p\oplus q_2)\\
        \vdots\\
        m\oplus T(p\oplus q_{n-4})
    \end{pmatrix}.
\end{equation*}

\begin{table}
    \centering
    \caption{The first three tweak constants on an Epyc 7251 processor. We denote the first one as $t_4$, since there are no dedicated constants for the least significant bits 3 to 0. This also implies that each tweak constant has a length of 16 bytes.}
    \label{tbl:tweak_constants}
    \begin{tabular}{c|c}
        $t_4$ & \texttt{82 25 38 38 82 25 38 38 82 25 38 38 82 25 38 38} \\ \hline
        $t_5$ & \texttt{ec 09 07 9c ec 09 07 9c ec 09 07 9c ec 09 07 9c} \\ \hline
        $t_6$ & \texttt{40 00 00 18 40 00 00 18 40 00 00 18 40 00 00 18}
    \end{tabular}
\end{table}
The first few constants are shown in Table \ref{tbl:tweak_constants}. Each constant consists of a repeating pattern of 4 bytes, thus reducing its entropy to at most 32 bits.

The tweak constants on our Epyc 7251 mostly equal those from~\cite{du2017secure}, who used a Ryzen 7 1700X. This suggests that AMD hardcoded these values, or at least uses a fixed seed to generate them on startup.
However, even fully randomizing these values on boot would not add any security, since they are shared across
\glspl{VM} and the hypervisor thus could easily compute them in advance, as shown above. 

We also performed these experiments on an AMD Ryzen 1950X, which only has \gls{SME} support.
On our first measurements we found that $t_8=t_9=0$, which led to repeating patterns within an encrypted page,
if the plaintext was all zeroes; some time later, after applying several operating system and BIOS updates, the tweak
values $t_8$ and $t_9$ changed, removing those patterns. This leads us to the conclusion that the tweak values
are influenced by firmware.

In summary, these results show, that \gls{XE} schemes in combination with missing integrity protection leak information about the tweak function. This is problematic especially in
the context of RAM encryption, where the tweak function is required to have low computational complexity.

\subsection{Updated XEX Encryption Mode}
\label{sec:encryption-mode-on-eypyc-embedded-3151}
We conducted the same experiments on an Epyc Embedded 3151 processor, which was released about 8 months after the Epyc 7251, and on an Epyc 7401P processor, which was released together with the Epyc 7251. On both processors, the system of linear equations did not have any solutions, i.e., AMD must have changed the encryption mode.

To reverse engineer the new encryption mode, we assumed that AMD did not greatly deviate from their previous implementations, and thus conducted a few experiments with slightly modified functions which used the same tweak values as before. This approach proved successful and yielded the new encryption function
\begin{equation*}
    \mathrm{Enc}_K(m,p)\coloneqq \mathrm{AES}_K\left(m\oplus T(p)\right)\oplus T(p),
\end{equation*}
and the matching decryption function
\begin{equation*}
    \mathrm{Dec}_K(c,p)\coloneqq \mathrm{AES}^{-1}_K\left(c\oplus T(p)\right)\oplus T(p).
\end{equation*}
As these equations show, AMD chose to use the \gls{XEX}~\cite{rogaway2004efficient} mode of operation, where a second tweak value is XORed to the AES encrypted ciphertext; in this case, both tweak values are identical.

The altered encryption function significantly complicates calculating the tweak constants, since simply decrypting a ciphertext at a different position does not yield usable results anymore:
\begin{align*}
    \mathrm{Dec}_K&\left(\mathrm{Enc}_K\left(m,p\right),q\right)\\
    &=\mathrm{AES}^{-1}_K\left(\mathrm{AES}_K\left(m\oplus T(p)\right)\oplus T(p)\oplus T(q)\right)\oplus T(q).
\end{align*}
Instead, the attacker needs to guess $T(p)\oplus T(q)$ and add this number to the ciphertext before decrypting. She can then check her guess by computing
\begin{align*}
    \mathrm{Dec}_K&\left(\mathrm{Enc}_K\left(m,p\right)\oplus T(p)\oplus T(q),q\right)\\
    &\overset{?}{=}\mathrm{AES}^{-1}_K\left(\mathrm{AES}_K\left(m\oplus T(p)\right)\right)\oplus T(q)\\
    &=m\oplus T(p)\oplus T(q).
\end{align*}
If all 128 bits of the tweak constants were chosen randomly, this operation would become infeasible; however, AMD still uses the repeated 4-byte pattern, so each tweak constant has only 32 bits of entropy.

Guessing these tweak constants is still computationally expensive, since one has to flush the respective TLB entry and the CPU caches when changing the encryption status of a page. We managed to partially work around this penalty by parallelizing our guesses, taking only around 30 minutes for each tweak constant. Given that even the newer CPUs still use the same tweak constants for every \gls{VM}, the hypervisor can pre-compute the table once in advance, so the slightly higher computation time becomes negligible in terms of security.

In summary, we showed that AMD implemented the well-known \gls{XEX} encryption mode. However, the tweak values have very low entropy and depend linearly on the physical memory addresses, enabling a malicious hypervisor to compute the entire table of tweak constants nevertheless. 
In the next two sections, we will exploit this fact and show how known plaintext can be used to place arbitrary code and data in the encrypted \gls{VM}.

\section{Cipher Block Moving Attack}
\label{sec:cipher-block-moving-attack}
As we have seen in the previous section, we can compute the tweak values for any physical address. In this section we show how a malicious hypervisor can use the knowledge of the tweak values together with known plaintext and missing integrity protection, to place 16-byte blocks containing some consecutive, controlled bytes.

This narrow attack vector already suffices to insert early returns in functions and skip parts of code, as shown in Section \ref{sec:case-studies}. In Section \ref{sec:arbitrary-code-exploit}, these byte sequences are exploited to build a full 16-byte encryption oracle, which allows us to execute arbitrary code on the highest privilege level within the \gls{VM}.

Contrary to previous work \cite{du2017secure,li2019exploiting}, which has used network I/O to create an encryption oracle, we do not need any control over the plaintext that gets loaded into the \gls{VM} in order to inject arbitrary data/code: Instead we simply use the plaintext that is already inside the \gls{VM} anyway.

\subsection{Attacker Model}
We assume that the attacker controls the hypervisor, which implies control over
the \glspl{NPT} and the ability to modify the \gls{VM}'s RAM. The attacker knows at least parts of the guest kernel's binary, which might be due to the unencrypted \texttt{/boot} partition or by using fingerprinting (see Section \ref{ssec:availability-of-known-plaintext}). We assume that the \gls{VM} is secured by \gls{SEV-ES}, implicating that the initial \gls{VM} image 
cannot be tampered with and the \gls{VMCB} is protected. 
We do not require, that the \gls{VM} communicates over the network or uses disk I/O.

\subsection{Tracking Guest Execution}\label{sec:tracking}
\begin{figure}
    \centering
    \includegraphics[width=0.5\textwidth]{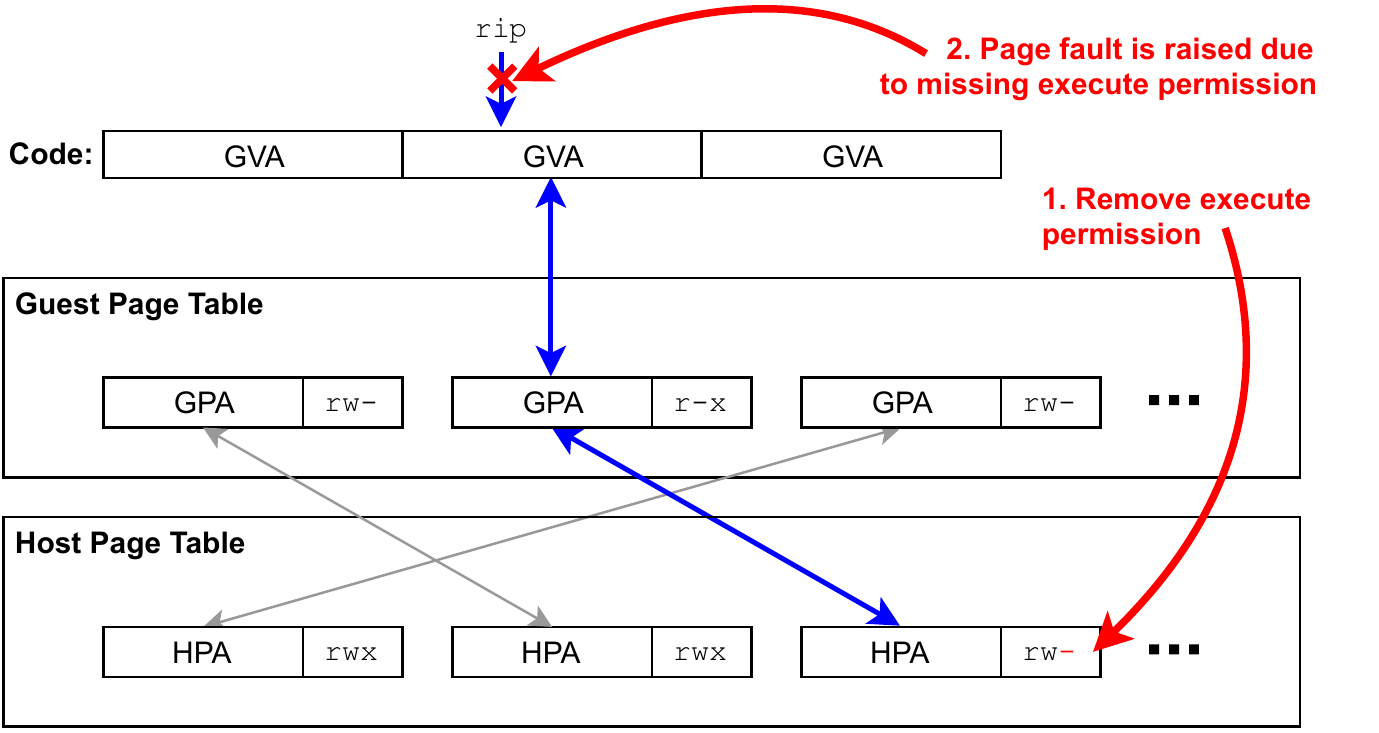}
    \caption{Page fault side channel. When the \gls{VM} tries to execute an instruction, the \gls{GVA} which the program counter (\texttt{rip}) is pointing to has to be resolved to a \gls{HPA}. This is accomplished by performing a walk through the \glspl{NPT}, while checking the respective permission flags. The hypervisor can force the \gls{VM} to page fault by removing the execute flag in the host page table entry. Subsequently, the hypervisor learns which page the \gls{VM} tried to execute. The same method can be used for detecting memory writes, by clearing the write flag instead.}
    \label{fig:pf-side-channel}
\end{figure}

To be able to make the \gls{VM} execute hypervisor-supplied code while being in a known state, we need to follow and eventually suspend its execution. We achieve this by using the page fault side channel, which has first been introduced in the context of Intel \gls{SGX}~\cite{xu2015controlled}. A schematic overview can be found in Figure~\ref{fig:pf-side-channel}. 
Since we control the hypervisor, and therefore the host page table, we can mark the relevant \gls{VM} pages as \emph{not writable} or \emph{not executable}. If the \gls{VM} then tries to issue a memory write or execute an instruction, a page fault is triggered and the corresponding hypervisor interrupt handler is called. 
The page fault exception contains the \gls{GPA} where the fault occurred.
The execution of the \gls{VM} can be resumed by marking the \gls{VM}'s
page as \emph{writable} respectively \emph{executable} within the host page table. 

Our attacks require computing tweak values, which in turn depend on \glspl{HPA}, so we have to infer the latter for
both the source and destination \glspl{GPA}. The \glspl{NPT} provide this translation.
Since we aim at injecting and executing code in the \gls{VM}, we need to find \glspl{GPA} that are mapped
as executable inside the guest. We cannot directly inspect the page
tables inside the \gls{VM}, but we can acquire this information by monitoring for
page faults due to missing execute permissions via the page fault side channel.

The guest kernel is a suitable target for code injection attacks, because it is executed with the highest
privileges
and is loaded to consecutive \glspl{GPA}. On modern Linux kernels, the base \gls{GPA}, to which the kernel gets loaded, is randomized by \gls{KASLR}. 
We present two methods for finding the guest kernel when \gls{KASLR} is active.
The Linux kernel is booted in two steps. First a small bootstrapper is loaded (to a fixed \gls{GPA}) that, amongst other setup tasks, is responsible for loading the actual kernel and for performing the \gls{KASLR}. In the first approach, we use our cipher block moving attack to modify the code of the bootstrapper, such that the \gls{KASLR} code is never executed. 
That way, the kernel is always loaded to a fixed \gls{GPA}. We present a more detailed description
in case study \ref{disable-kaslr}.
For the second approach, we monitor the naturally occurring page faults during \gls{VM} startup.
We exploit, that the kernel is loaded to continuous \glspl{GPA} and that the memory accesses before loading the kernel to a randomized address, are quite deterministic. This allows us to identify the memory accesses related to loading the kernel, which gives us the \glspl{GPA} of the kernel.

At the moment, AMD's patched Linux kernel cannot be complied to run as a SEV-ES guest and use \gls{KASLR} at the same time, as the compile time options for these features exclude each other (see CONFIG\_SEV\_ES\_GUEST and CONFIG\_RANDOMIZE\_BASE in arch/x86/Kconfig in AMD's kernel repository~\cite{amdSEVESKernelRepo}). We are not aware of any fundamental conflict between these two features and thus suspect that this is only a temporary implementation issue.

We tested these methods on SEV secured machines, but as they do not rely on an unencrypted \gls{VMCB} they should also work with SEV-ES.
As mentioned in the
attacker model, the kernel code can be assumed to be entirely known to the attacker and thus serves as a reliable source
for ciphertext blocks with known plaintext, which can be copied to other places in the kernel to trigger malicious
behavior.

\subsection{Placing partially controlled Plaintext}
Knowing the destination address in \gls{VM} memory we can now start to construct our attack primitive. Since \gls{SEV} lacks any integrity protection, the hypervisor can modify the contents of the entire guest's memory. Randomly guessing ciphertexts is rather unlikely to yield meaningful plaintext and will, especially in the case of code, most probably crash the \gls{VM}. However, since we can compute the tweak values for any given address, we can re-use existing ciphertext blocks after applying slight adjustments.

We assume that we want to place a 16-byte block $m$ at address $p$. We then need to find an address $q$ holding a known 16-byte plaintext block $m'$, which satisfies the following property:
\begin{alignat*}{2}
	&&m\oplus T(p)&=m'\oplus T(q)\\
	&\Leftrightarrow\;&m'&=m\oplus T(p)\oplus T(q)
\end{alignat*}
Copying the corresponding ciphertext block from $q$ to $p$ and decrypting it, yields the desired plaintext block $m$:
\begin{align*}
	\mathrm{Dec}_K&\left(\mathrm{Enc}_K\left(m\oplus T(p)\oplus T(q),q\right),p\right)\\
	&=\mathrm{AES}^{-1}_K\left(\mathrm{AES}_K\left(m\oplus T(p)\oplus T(q)\oplus T(q)\right)\right)\oplus T(p)\\
	&=\left(m\oplus T(p)\oplus T(q)\oplus T(q)\right)\oplus T(p)\\
	&=m.
\end{align*}

To target the \gls{XEX} encryption mode the copied ciphertext block needs to be slightly adjusted, by adding $T(p)\oplus T(q)$:
\begin{align*}
	\mathrm{Enc}_K&\left(m\oplus T(p)\oplus T(q)\right),q)\oplus T(p)\oplus T(q)\\
	&=\mathrm{AES}_K\left(m\oplus T(p)\right)\oplus T(p).
\end{align*}
Decrypting this at address $p$ will then yield $m$.

The complexity of the bit sequences a malicious hypervisor is able to create with this method is limited by several factors. The first is the diversity of the known plaintext blocks, i.e., whether they have enough entropy. The next limitation is the 32-bit periodicity of the tweak values (which we can control by
choosing the \gls{HPA} a \gls{GPA} gets mapped to), so we can expect to be able to control at most 4 bytes of any 16-byte block in a reliable way. Finally, for our processors we found that only 28 tweak constants are linear independent, so for each guest page the hypervisor can choose from up to $2^{28}$ different base addresses which yield different ciphertext blocks. This suggests a rough upper bound of 3 bytes per block, which an attacker is likely to be able to fully control, if given enough plaintext and memory.

In our experiments, we found that we can very reliably find a fitting pair $m'$/$q$ for any sequence of two bytes, given about $8$ MB of known plaintext. We copied the \texttt{.text} (code) section of the Linux kernel bootstrapper as it gets loaded into memory, which can be easily located due to the lack of randomization of its load \gls{GPA}. In addition, we can use the \texttt{.text} section of the kernel binary itself, after locating it in memory with one of the previously described methods.

\subsection{Code Injection}\label{sec:code-injection-primitive}
We now show how the two controlled bytes per block can be used to modify existing \gls{VM} code, allowing us to redirect the control flow and to insert arbitrary 2-byte instructions.

\begin{figure*}[t]
    \includegraphics[width=\textwidth]{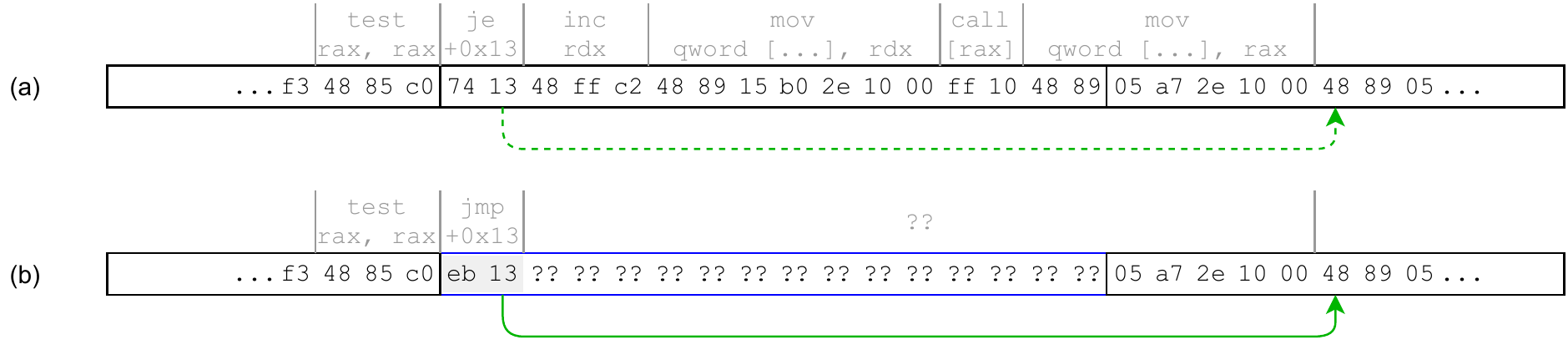}
    \caption{Example for changing execution flow by replacing one 16-byte block of code. While the base program (a) branches conditionally depending on the value of the \texttt{rax} register, the patched version (b) has this branch replaced by an unconditional one. The remainder of the inserted block consists of uncontrolled bytes, which are not expected to form a sequence of meaningful instructions.}
    \label{fig:single_block_move}
\end{figure*}
Instructions on x86-64 have variable length and might share prefixes, so we have to consider whether we change or break an existing instruction when injecting our 16-byte block. Also we have to ensure that the uncontrolled bytes of our block do not get executed. The easiest way to achieve this is by finding a 16-byte aligned instruction and overwriting it with a short branch instruction, like \texttt{ret} or \texttt{jmp} (Figure \ref{fig:single_block_move}). This simple modification already suffices to completely disable \gls{KASLR} (see Section \ref{disable-kaslr}).

Finding a 16-byte aligned instruction for an injection point is rather easy for 64-bit code: For performance reasons, most compilers align functions and frequently used chunks of functions to an architecture-specific value, which usually happens to be 8 bytes on x86-64, so we can expect around every second function to be aligned to a 16-byte boundary. 

To avoid executing the uncontrolled bytes of a block, we always have to insert a jump instruction -- which takes both usable bytes of a block, so this method only allows us to skip small parts of the underlying code.

\begin{figure}[t]
    \centering
    \includegraphics[width=0.48\textwidth]{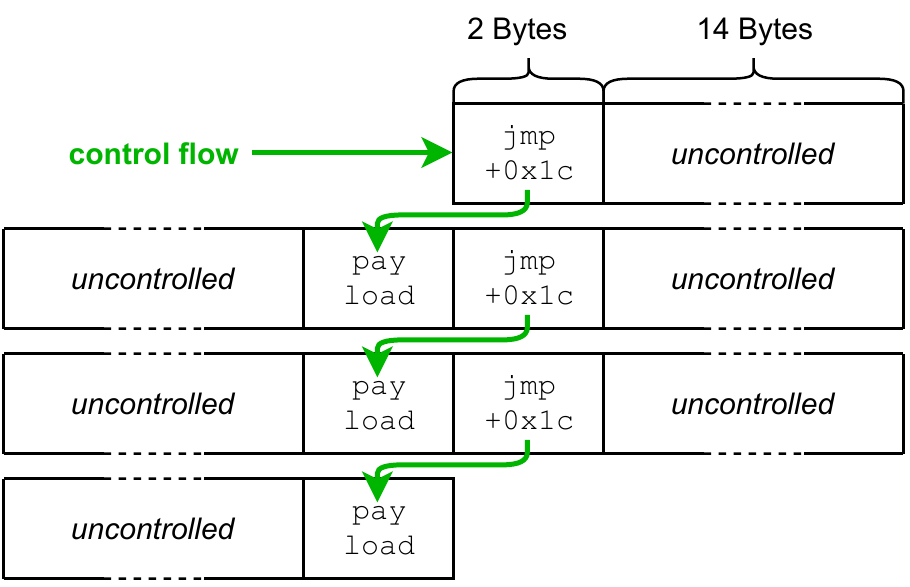}
    \caption{A sequence of consecutive 16-byte blocks blocks are chained together to get small contiguous chunks of code, which are connected by unconditional 2-byte \texttt{jmp} instructions to avoid executing the uncontrolled bytes in between. Thus two bytes of every second block can be used to execute arbitrary 1-byte or 2-byte instructions (payload). The last payload may either redirect to original code (e.g., by returning), or enter a loop.}
    \label{fig:complex-injection}
\end{figure}

To insert other instructions, we propose the layout shown in Figure \ref{fig:complex-injection}. First we inject a \texttt{jmp} at a 16-byte aligned instruction. With this, we jump to offset 14 of the following block, where we can
place an arbitrary two byte instruction (payload). Then we can use the first two bytes of the following block to again jump 
to the next payload location. This way we maximize the amount of consecutive bytes that we can control.

In Section \ref{sec:case-studies} we successfully use this method to disable \gls{KASLR} and illustrate a fast \texttt{cpuid}-based 16-byte encryption oracle. Finally, in Section \ref{sec:arbitrary-code-exploit}, we build another 16-byte encryption oracle which solely relies on ciphertext block moving and the ability to provoke a context switch between hypervisor and \gls{VM} at a precise moment in time. We demonstrate how the latter can be achieved by using emulated instructions or page faults. In contrast to the \texttt{cpuid}-based 16-byte encryption oracle, this final encryption oracle does not depend on the capability of the hypervisor to modify the result of emulated operations, so it is difficult to mitigate without introducing proper integrity protection. Both encryption oracles allow us to execute arbitrary code within the \gls{VM}.

\section{Attack Case Studies} \label{sec:case-studies}

\subsection{Control Flow Modification for KASLR} \label{disable-kaslr}
To be able to track \gls{VM} execution, the hypervisor needs to know the base \gls{GPA} of the kernel, which is randomized by \gls{KASLR}. In order to perform a first demonstration of our attack primitives, we disable \gls{KASLR} using the one-block code injection method from Section~\ref{sec:code-injection-primitive}, effectively placing the kernel at a well-known, constant \gls{GPA}.

When loading the kernel, the bootstrapper code calls the function \Func{void choose\_random\_location(...)}, which is defined in \Path{/boot/x86/compressed/kaslr.c}. The function checks whether the user provided the \textit{nokaslr} command line option; if this is the case, it returns immediately. Else the function computes random physical and virtual base addresses for the kernel and writes them into the supplied pointer arguments. So, to disable \gls{KASLR}, it is sufficient to place a \texttt{ret} at an early location in the function.

To find the right point in time to modify \Func{choose\_random\_location}, we utilize the page fault side-channel as explained in Section \ref{sec:tracking}. We remove the write permissions to the physical page \emph{after} the first part of the targeted function is copied.
This causes a page fault to be triggered as soon as the boot loader is done copying the first part of the function and tries to copy the next one. When handling the page fault, the hypervisor places the block containing the \texttt{ret} instruction, and then resumes execution. In our experiments, the targeted function was always located near
the end of the bootstrapper's \texttt{.text} section, which means that we already have a few MB of known plaintext
at this point, depending on the kernel binary. In addition, the \texttt{ret} instruction only requires a 1-byte opcode, which greatly reduces the amount of known plaintext that is needed to inject the instruction.
As stated above, AMD's patched Linux kernel can currently not be configured to both use \gls{KASLR} and run as an SEV-ES guest. Thus we only tested this attack on SEV secured \glspl{VM} (without the ES extension).

\subsection{Using CPUID as an Encryption Oracle} \label{sec:using-cpuid-as-an-encryption-oracle}
Our code injection primitive can be combined with the hypervisor-emulated (intercepted) \texttt{cpuid} instruction to gain control over certain general purpose registers and build a high performance 16-byte encryption oracle.

As explained in \ref{ssec:amd-memory-encryption}, the content of the \gls{VMCB} gets encrypted and integrity protected upon a \texttt{\#VMEXIT} in case \gls{SEV-ES} is enabled. 
This prevents a malicious hypervisor from manipulating its content; however, in order to emulate instructions like \texttt{cpuid}, the value of certain registers is still shared via the \gls{GHCB}. While the guest owner may disable instruction emulation, they are an important virtualization feature that allows for fine-grained control over exposed hardware features
as well as keeping the \gls{VM}'s environment consistent in case of live migration.

\gls{OS} kernels frequently call the \texttt{cpuid} instruction during startup to retrieve information about the system's capabilities and topology. 
Since the results of these calls often get directly stored in memory for caching purposes (e.g., the vendor string), this poses an easy target for injection attacks.

First we determine the \gls{HPA} of the \texttt{cpuid} call and the associated memory store; then we inject a block 
containing an unconditional jump to the \texttt{cpuid} instruction after the memory store in order to create a loop. On
each \texttt{cpuid} call, the hypervisor sets the return registers, resumes execution and waits for the next 
\texttt{cpuid} call. When this call occurs, the data from the last call has been stored, so the hypervisor can copy the 
encrypted data to the desired location.

We implemented this exploit in the \texttt{get\_model\_name} function (\texttt{arch/x86/kernel/cpu/common.c}) of the Linux kernel, since it writes the \texttt{cpuid} result to a contiguous block of 16 bytes, which can be directly used as an encryption oracle. 
One could also use our basic injection attack to create such a \texttt{cpuid} loop. The results
can be stored on the program's stack memory, whose \gls{GPA} can be determined by the stack detect gadget which will
be presented in the next section.

Since we only need one \gls{VM}/hypervisor context switch per 16-byte block, this channel is very efficient: We encrypted 1'000'000 blocks (16 MB) within around 37.5 seconds, suggesting a bandwidth of around 3.41 MBit/s or 426.67 KB/s.

\section{Executing arbitrary Code}\label{sec:arbitrary-code-exploit}
The previous two examples have shown that even little modifications of control flow can have a severe effect on the system's overall security.
However, our ultimate goal is to execute arbitrary code, without having to rely on the ability to control register contents through an intercepted instruction, or use of I/O.
We will advance the 4-byte block chaining method from
\ref{sec:code-injection-primitive}, to inject a program into the \gls{VM}, which writes arbitrary data into a 16-byte block of memory. This block encryption oracle enables us to execute arbitrary code with kernel privileges inside
the \gls{VM}. We show that the oracle can be easily used to construct a decryption oracle as well.

The basic idea is to inject a small code gadget into the \gls{VM}, that performs some computations in order to write 4 bytes of plaintext into a 32-bit register. Next we push this register onto the stack, to get an encrypted version of our plaintext; this serves as an intermediate 4-byte encryption oracle, so we are able to control $2+4=6$ consecutive bytes. We then use this increased payload size to repeat the same process with 64-bit registers, finally giving us control over the full 16 bytes of a block.

\subsection{Triggering the Hypervisor} \label{triggering-the-hypervisor}
The proposed attack needs careful synchronization between \gls{VM} and hypervisor, such that the hypervisor can suspend execution at a precise point in time and modify guest memory. We propose two different mechanisms to achieve this.
The first mechanism utilizes the \texttt{cpuid} instruction, which is emulated by the hypervisor and features a 2-byte opcode: Each time \texttt{cpuid} is executed, the hypervisor is called to emulate it. So, by interleaving the injected instructions with \texttt{cpuid} calls, we can precisely redirect execution to the hypervisor.

The \texttt{cpuid} calls clobber the \texttt{eax}, \texttt{ebx}, \texttt{ecx} and \texttt{edx} general purpose registers, so they are not usable for the constructed gadgets. Also, the \texttt{eax} register (which determines the requested leaf ID) should be cleared beforehand to avoid calling additional handling logic in the hypervisor -- leaf 0 just returns the vendor ID.

It is convenient to use the \texttt{cpuid} instruction because it has a simple handler in the KVM hypervisor.
Instead of \texttt{cpuid}, we could also use other instructions that are intercepted by the hypervisor and require at most a 2-byte opcode, like \texttt{rdtsc} (for a complete list see~\cite{AMD2019architecture}).
As stated in \ref{sec:interception}, instruction interception is important, as it, for example, allows
fine grained control over exposed hardware features as well as live migration.
However, as stated in \ref{sec:interception}, the guest owner can choose to disable instruction interception for the \gls{VM}, with the downside of losing the mentioned functionality.

A more complex alternative to the \texttt{cpuid}-based execution transfer is the usage of the page fault side channel, which also allows a precise interruption of the \gls{VM}.
If we want to interrupt the \gls{VM} between two injected instructions, we ensure that they reside on two different pages $p_1$ and $p_2$, and remove the execute permission in the hypervisor's \gls{NPT}. This way, the \gls{VM} gets interrupted before the first instruction in $p_2$ gets executed. We then remove the execute permission for $p_1$, such that the hypervisor gets triggered another time to remove execute permissions for $p_2$ once again.

In the following we will use \texttt{<sync>} to express that one of the just described mechanisms must
be used, to interrupt the \gls{VM} at a certain point in time.

\subsection{Finding the Stack} \label{subsec:finnding-the-stack}
In order to use the stack for our encryption oracle, we need to get the \gls{HPA} of the related stack pages. We solve this problem by combining the synchronisation mechanism with the page fault side channel.

We use the attack primitive from Section \ref{sec:code-injection-primitive} to construct an instruction sequence \texttt{<sync>; push rdi; <sync>}. \texttt{<sync>} triggers the hypervisor, which then removes write access from all memory pages belonging to the \gls{VM}, and resumes execution. The following \texttt{push rdi} tries to write to the non-writable stack memory page, and subsequently raises a page fault in the hypervisor. The page fault exception information yields the corresponding 
\gls{GPA} and thus the \gls{HPA} of the stack. However, on SEV-ES the page offset is masked out.
To overcome this, we take a copy of the whole page, while the hypervisor is handling the page fault
and compare it with the content of the page at the second \texttt{<sync>}. The position of the
ciphertext block that was changed by the \texttt{push rdi} operation gives us the exact offset of the stack inside the page.

If the write address of the \texttt{push rdi} is near the end of a page, the hypervisor may issue an extra \texttt{pop rdi} instruction to ensure that the next stack operation writes to the same page. This also significantly eases restoring original execution after inserting the encryption oracle code.

\subsection{4-Byte Encryption Oracle}
Originally, x86 only supported 16-bit and 32-bit operands. When the CPU vendors implemented support for native 64-bit operations, they did not add new opcodes for every general purpose instruction (e.g., arithmetic and memory-to-register/register-to-memory); instead, they introduced the \emph{REX} prefix, which, when put before an instruction's opcode, upgrades its operands to 64-bit mode. Since in 32-bit mode most general purpose instructions are encoded using at least 2 bytes, this prefix extends them to 3 bytes -- but our attack primitive only supports 2 bytes of payload. However, when adding 64-bit support, the 1-byte \texttt{push \emph{reg}} instructions were redefined to only support 64-bit registers, so we can use the payload to perform stack writes. We thus can use 32-bit instructions to control the lower half of some registers, and then push those onto the stack. Hence we can control the lower 4 bytes of a 16-byte block, so the possible payload is doubled, enabling us to use 64-bit instructions for the next step.

\begin{figure}[t]
    \centering
    \includegraphics[width=0.48\textwidth]{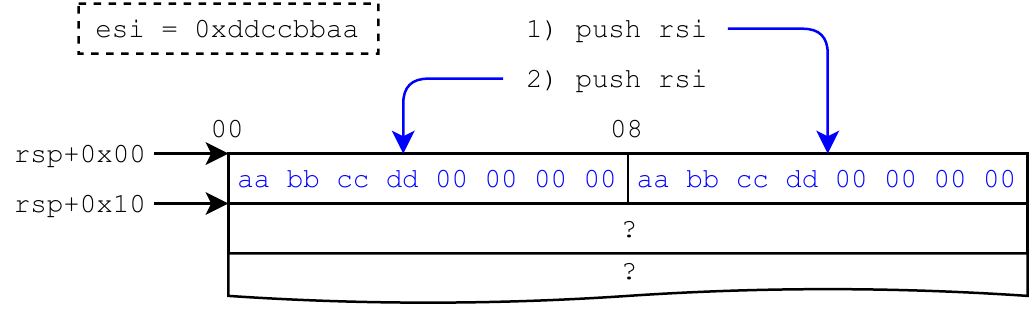}
    \caption{Layout of stack after pushing the 32-bit \texttt{esi} register. The stack pointer is decreased when pushing a register, so, depending on the stack pointer's original alignment, we might have to push the register another time to set the lower address part of a 16-byte block. Since this is a 64-bit operation, the (zeroed) higher 32-bits of \texttt{rsi} are pushed as well. Due to the endianness of x86 the lower significant bytes end up first, the higher bytes last. We thus finally get a 16-byte block where we control the first 4 bytes.}
    \label{fig:4-bytes-stack}
\end{figure}

x86 is a little endian system, so when we push a register to the stack, its bytes are stored in reversed order. This means, if we set the least significant 32 bits of a register and push it to the stack, those bits will be placed at lower addresses (Figure \ref{fig:4-bytes-stack}). If the stack pointer has been 16-byte aligned before our first push, the controlled bytes will then reside in the middle of the 16-byte block, where we cannot chain them with another block. So we have to push the register \emph{a second time} -- now the stack pointer is 16-byte aligned, and the payload resides at the block beginning. Depending on the stack page offset and the amount of blocks being created, one might have to add some \texttt{pop} instructions to free up stack space before proceeding with the next block.

As a last building block, we need a gadget to place an arbitrary 32-bit value into a register. This gadget can be constructed via a simple combination of increments and left shifts: First, the register is cleared by XORing it with itself (this also automatically clears the upper 32-bit of the corresponding 64-bit register). To add a 0 bit, the register is just shifted; to add a 1 bit, the register is incremented and then shifted. This will take at most 31 rounds until the most-significant bit has been set. All the involved instructions have 2-byte opcodes.
\begin{figure*}[t]
    \centering
    \includegraphics[width=0.9\textwidth]{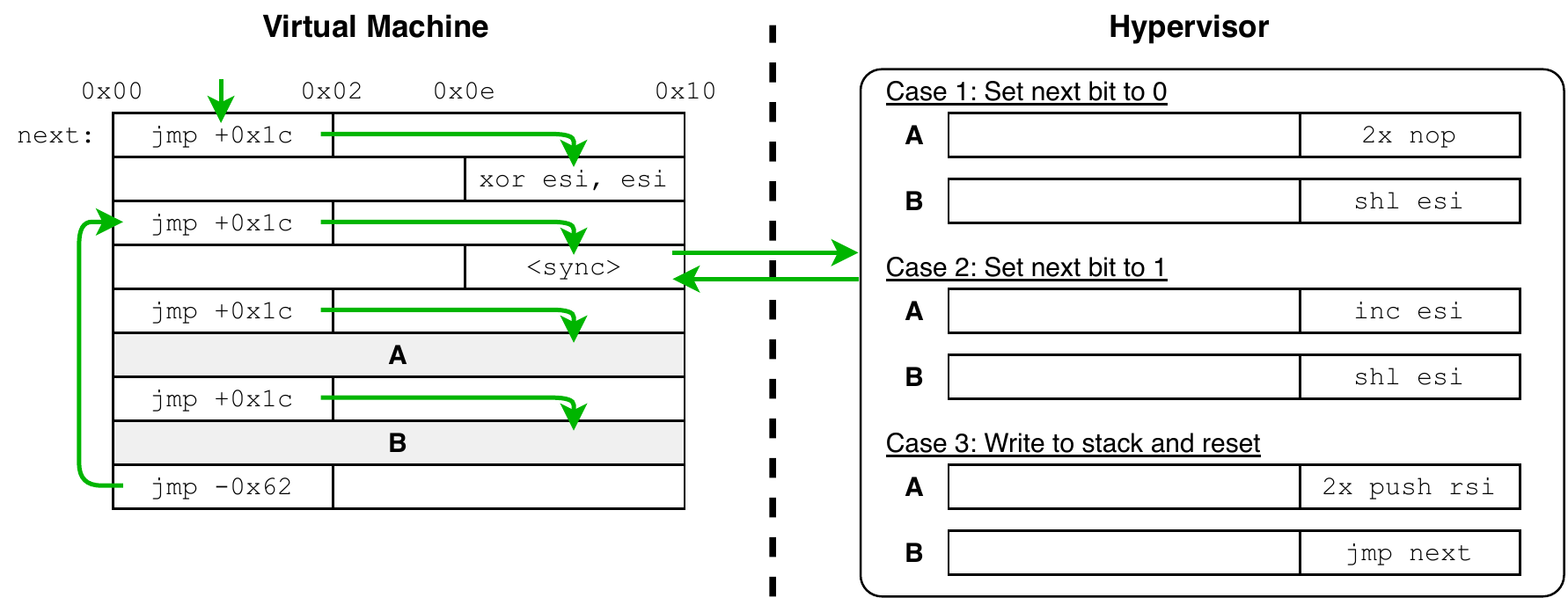}
    \caption{Schematic of the 4-byte encryption oracle. Each row represents a 16-byte block (not to scale), control flow jumps are denoted by arrows. On each use of the sync mechanism from Section \ref{triggering-the-hypervisor} (for example a \texttt{cpuid} call), the hypervisor replaces blocks A and B depending on the desired action: It may either shift 0s and 1s into \texttt{esi}, or push the \texttt{rsi} register two times to the stack to get an encrypted 16-byte block. This process can be repeated arbitrarily often.}
    \label{fig:32-bit-oracle}
\end{figure*}
The final block layout forming the 32-bit oracle is shown in Figure~\ref{fig:32-bit-oracle}.
To move the payload from the location where it gets encrypted
to another memory location, we need to consider the XOR difference of the tweaks used at these two memory locations
and XOR it with our payload before using the 4-byte oracle.

In summary, we are now able to control $4$ bytes per 16-byte block. In the next paragraph,
we show that this is sufficient to inject a program allowing us to control a whole $16$ byte block.

\subsection{16-Byte Encryption Oracle}
The 16-byte encryption oracle works very similar to the 4-byte encryption oracle. First, we ensure that the stack is
16-byte aligned; if we used the described process for creating the 4-byte encryption oracle, we already have this
information. Then we use the same strategy as in the 4-byte encryption oracle to load the two 64-bit chunks of our
plaintext into 64-bit registers and push them onto the stack. Since we made sure that the stack was 16-byte aligned
before the first push operation, we now have an entire 16-byte aligned 16-byte block in memory, which only needs to be
copied to the desired location.

The formerly introduced 4-byte oracle allows us to use 6-byte instruction gadgets, so after subtracting the necessary \texttt{jmp} instructions we can use 4 bytes of payload. This is sufficient for most 64-bit register-to-register arithmetic. Though there might be more efficient methods for assigning hypervisor-defined values to a 64-bit register, we reuse the increment/shift method for sake of simplicity.

\begin{figure}
    \centering
    \includegraphics[width=0.48\textwidth]{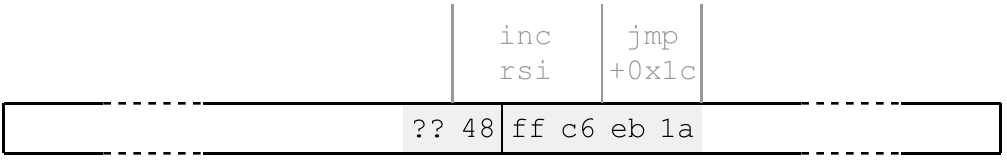}
    \caption{Example for injection of a 3-byte opcode payload followed by an unconditional jump, using a block created with the cipher block moving primitive, and one block from the 4-byte encryption oracle. It is desirable to fully use the 4 bytes from the encryption oracle, since finding a fitting block for the cipher block moving primitive requires more complexity, when the number of payload bytes increases.}
    \label{fig:long-opcode-format}
\end{figure}

The implementation is very similar to the 4-byte oracle: All instructions involving the target register (\texttt{rsi}) are extended to 64-bit using the REX opcode prefix. Additionally, instead of pushing \texttt{rsi} twice, it is only pushed once and another iteration is started to push another value. This way, we can fully control all 128 bits of the plaintext block. Figure \ref{fig:long-opcode-format} shows an excerpt of a gadget using a 3-byte opcode payload.

In summary, we are able to encrypt arbitrary 16-byte values, by injecting a program into the \gls{VM} that
performs some computations in order to write data into encrypted memory owned by the \gls{VM}.

\subsection{Code Execution allows stealthy Decryption} \label{sec:encryption-implies-decryption}
Throughout this section we have shown how to execute arbitrary code via a self-bootstrapping, non I/O dependent encryption oracle. This of course raises the question if it is possible
to create a decryption oracle, with a similarly low set of requirements.
We now show how a decryption oracle can be constructed by extending an idea of Hetzelt and Buhren \cite{hetzelt2017security}.

As explained in Subsection \ref{ssec:amd-memory-encryption}, the encryption
status of a page can be controlled via the C-bit, in each page table entry.
This allows the \gls{VM} to share pages with the hypervisor.
Hetzelt and Buhren show that using an encryption as well as an decryption oracle, the hypervisor
can insert a shared page into the page table of a process running inside the VM.
The hypervisor can then copy the content of an encrypted page into the shared page.
In their approach, they use a decryption oracle in order to find a free entry
in the page table of a victim process running in the VM.
We do not need a decryption oracle for this approach:  Allocating a shared page, as well as
copying some data to it, can be done via an injected program instead.

Thus, we conclude that the existence of an encryption 
oracle immediately implies a decryption oracle.
Furthermore this method is very stealthy compared to using loggable network communication to extract data,
like in \cite{morbitzer2018severed,li2019exploiting}. In addition this allows for very high throughput, as
the copy rate of the injected program is only limited by the VM's ability to write to RAM.

\section{Comparison to Related Work } \label{sec:comparison-and-evaluation}
First, we present a performance analysis of our 16 byte encryption oracle, before
comparing it to encryption (and decryption) oracles constructed in related work.

\smallskip\noindent\textbf{Throughput of our oracle}
We performed our experiments on two different setups.
Initially we used an AMD Epyc 3151 CPU with 16 GB of RAM. The
host was running Ubuntu 19.04 with Linux kernel version 5.0.18
and the guest was running Ubuntu 19.04 with kernel 5.0.0-27-generic with 1 GB of RAM.
The used QEMU version was 2.12.0.
As the BIOS on this machine does not yet support \gls{SEV-ES}, we also used a machine with an Epyc 7401P processor, that provided the necessary BIOS support, to verify our results under SEV-ES. As mentioned in the introduction, SEV-ES needs extensive software support in the Linux
kernel, the QEMU emulator and the UEFI of the \gls{VM}.
We used the versions from AMD's official repositories~\cite{amdSEVESKernelRepo,amdOvmfRepo,amdQemuRepo}.
We made use of the SEVered framework \cite{severedframework} to inject page faults into the \gls{VM}.
    
To evaluate the performance of our encryption oracle, we set up a program that waits 
for a trigger before calling the function in which we injected our gadgets.
First, we bootstrap the 16-byte encryption oracle via the stack detect gadget and the 4-byte encryption
oracle. Then we use it a thousand times to encrypt 16 bytes of payload data. On our unoptimized prototype the setup part takes
$0.62$ seconds and the payload encryption needed $75.86$ seconds. This
translates to a throughput of $211$ Bytes per second for the 16 byte oracle.

Our prototype implementation focuses on ease of implementation and debugability, thus the performance can be improved by writing more than one bit to the \texttt{rsi} register before interrupting the computation via the sync mechanism. A sequence of zeroes could be written by inserting an $x$-bit left shift (4 byte opcode), instead of performing $x$ rounds with a single bit left shift.
Furthermore we could simply increase the number of instructions we execute each round, to decrease
the number of interrupts/context switches and write operations which require expensive flushes.

\smallskip\noindent\textbf{Comparison}
We compare our results to encryption/decryption oracles constructed in related work.
If not stated otherwise all attacks assume a malicious hypervisor.
An overview can be found in Table~\ref{tbl:we-are-the-best}.

\begin{table}
    \centering
    \caption{Comparison of different approaches for encryption oracles. $^{1}$Li et al.~\cite{li2019exploiting} only specify the decryption rate, but it should be similar to the encryption rate.}
    \label{tbl:we-are-the-best}
    \begin{tabular}{M{21mm}|M{11mm}|M{11mm}|M{11mm}|M{11mm}}
        & Du et al.~\cite{du2017secure} & Li et al.~\cite{li2019exploiting} & \texttt{cpuid} & Cipher Block Moving \\ \hline
        Needs service in VM & yes & no & no & no \\ \hline
        Relies on I/O & yes & yes & no & no \\ \hline
        Needs instruction emulation & no & no & yes & no \\ \hline
        Encryption rate (B/s) & unknown & 200$^{1}$ & 426670  & 211 
    \end{tabular}
\end{table}

\textbf{Du et al.}~\cite{du2017secure} were, to the best of our knowledge, the
first to discover the original
encryption mode of \gls{SME} as well as the tweak values. 
Their experiments were performed on an AMD Ryzen CPU without \gls{SEV} support (AMD Epyc 7xx1
CPUs were not readily available at that time).
They constructed an encryption oracle for a self-built simulation of \gls{SEV}. Their attack
requires knowledge of the tweak values, an Nginx server running in the \gls{VM} and is not mitigated
by \gls{SEV-ES}. 

They found that Nginx stores parts
of the data sent to it in consecutive 16 byte blocks at fixed offsets inside a page.
Building on this, they send an HTTP packet whose payload is designed in a way, that the parts going
to these offsets contain exactly the tweak values of said offsets.
This way, the data encrypts to a constant ciphertext, making it easily detectable in a memory dump.

They use this to encrypt code and execute it in the \gls{VM}.
In contrast to our encryption oracle they rely on self-generated network traffic getting processed by
an Nginx webserver inside the \gls{VM} as well
as the discussed memory management behavior of Nginx.
It is unclear whether different services, or even different versions of Nginx, show a similar exploitable behavior.
They do not give performance measures.

\textbf{Li et al.} \cite{li2019exploiting} showed how to create an encryption/decryption oracle by
leveraging unprotected DMA operations, knowledge of the tweak function and control over the NPTs.
For the demonstrated attack, they also require network traffic, whose frequency linearly scales with
the throughput of their oracles. Their attack works with \gls{SEV-ES}.

According to them, DMA is the most common method
used by \glspl{VM} to perform I/O Operations. They exploit that current IOMMU hardware (which is
responsible for performing DMA) only supports
one memory encryption key, while \gls{SEV} uses one key for the hypervisor as well as an additional key per \gls{VM}.
Thus all DMA operations must be performed on memory pages $p_s$ that are shared between the hypervisor and
the \gls{VM}, i.e., encrypted with the hypervisor's encryption key. This means if the guest wants to write
data via DMA, it first needs to prepare the content in a private page $p_p$ 
before copying the
content into $p_s$. Reading data
via DMA works the other way around.

The general idea for their decryption oracle is to manipulate the content of $p_p$,
before its content is copied to $p_s$.
For their decryption oracle they use DMA write operations.
To decrypt the memory at address $q$ they copy it 
into $p_p$, before it gets copied to $p_s$. 
In order to get the \gls{GPA} of $p_p$ they use the page fault side channel.
They demonstrated their ideas based on DMA operations related to OpenSSH network traffic.

For the decryption oracle they are limited to the packets sent by the \gls{VM}. 
Furthermore, they show that they can make their oracle harder to detect
by only overwriting parts of $p_p$ that contain known metadata spanning at least
a whole 16 byte aligned block.
This way, they can restore the overwritten parts before sending the package over the network.
Assuming a packet rate of 10 packets per second they showed that their decryption oracle has
a throughput of about $200$ B/s.

For the encryption oracle they
can use self-generated packets, even if there is no service listening.
The \gls{VM} can however observe the amount of dropped packages.
They did not give any data for the throughput of the encryption oracle.
But since the construction is similar to the decryption case,  
its throughput should scale in a comparable manner with the packet rate.

For the encryption oracle they do not state whether the idea of replacing the payload
with known metadata can be applied.
If this is not possible, the \gls{VM} can observe the packages that get destroyed by the encryption
oracle. Our encryption oracle is not affected by such problems, because we take over 
the control flow that processes the data, instead of trying to manipulate data used by
the regular control flow.

Like we have shown above, our encryption oracle reaches a slightly higher throughput with our
prototype implementation, although they based their measurements on a SSH packet rate of $10$ pps,
which is quite high for user generated input (one packet roughly equals one keystroke).
Since we do not depend on I/O, we can achieve our throughput independently of the rate of network packages.
While they claim that their approach can be applied to any DMA I/O performed by the \gls{VM}, it is
unclear which of them 
sport known metadata that spans at least a 16-byte aligned memory block in order to make the attack stealthy.

\section{Countermeasures} \label{sec:countermeasures}
    Our code injection attacks, as well as the injected 16-byte encryption oracle,
    build on the missing integrity protection, the reverse engineered
    tweak values, known plaintext and the page fault side channel.
    The high performance \texttt{cpuid} encryption oracle 
    from the case study \ref{sec:using-cpuid-as-an-encryption-oracle} also requires that 
    the \texttt{cpuid} instruction is interceptable by the hypervisor.
    In the following paragraphs we discuss how changes in these areas
    influence our attack.
    
    \subsection{Integrity Protection}
    With cryptographic integrity protection, the encryption system could detect blocks
    created with the cipher block moving approach. This would prevent us from
    injecting code/data into the \gls{VM}, mitigating the attacks presented in this paper, as well as
    all of the related work mentioned in Section \ref{sec:comparison-and-evaluation} 
    with the exception of the application fingerprinting
    presented in \cite{werner2019severest} and the attacks on the AMD \gls{SP} from \cite{buhren2019insecure}.
    In January 2020, AMD released a whitepaper on a planned future extension
    called \gls{SEV-SNP}~\cite{sev-snp}. Instead of adding strong, cryptographic integrity protection,
    they
    propose an access right based system called \gls{RMP}, that assigns each physical memory page either to the hypervisor, a specific \gls{VM} or to the \gls{SP}. Only the owner of a page is given
    write access. The \gls{RMP} will be manageable by the hypervisor via an instruction set extension.
    For \gls{SEV-SNP} secured \glspl{VM}, the \gls{RMP} also contains the supposed \gls{GPA} of the page inside the \gls{VM} as well as a "validated" flag, that is always false for new \gls{RMP} entries. The "validated" flag can only be manipulated by the
    \gls{VM} that the page is assigned to. This way, AMD intends to prevent remapping attacks like the one in~\cite{morbitzer2018severed}, as they would require the \gls{VM} to validate multiple
    pages for one \gls{GPA}.
    While this mechanism is not able to detect that a cipher text block was manipulated, the lack of write
    access to a page would prevent us from performing our cipher block moving attack.
    However, given the required architectural changes, it is not foreseeable when \gls{SEV-SNP} will be
    available.

    \subsection{Tweak Function}
    Without the knowledge of the tweak values we could no longer predict the effect of a
    cipher block move.
    \cite{li2019exploiting} claims that "Future versions of the tweak function will be 
    implemented as $T(k,a)$ where $a$ is the physical address and $k$ is a random input that
    changes after every
    systems boot". For the non \gls{XEX} version of the encryption scheme, considered by them,
    this would not make any difference, since our method from
    Section \ref{sec:reversing-the-encryption-mode} can be implemented in a kernel
    module to recalculate the tweak values at run time, with very little overhead.
    For the \gls{XEX} version, discovered by us, we demonstrated in Subsection \ref{sec:encryption-mode-on-eypyc-embedded-3151}
    how to brute force the tweak values at run time, as long as they stay 32 bit 
    periodic (or similarly low periodicity). While the tweak recovery process takes about $30$ minutes per tweak, we want to stress that the decision to reboot is under the control of the malicious hypervisor.
    However, we are unaware of any method to directly calculate the tweak values, like it
    was possible with the previous version.
    We believe that using 128-bit randomized tweak values are a mitigation to this attack vector.
    
    \subsection{Fixing the Page Fault Side Channel} \label{ssec:fixing-the-page-fault-side-channel}
    In our opinion, completely removing the hypervisor's ability to observe the page faults of the \gls{VM} is not realistic, since the hypervisor needs this information for memory management purposes.
    However, we believe that the amount of leaked information can be reduced,
    by restricting the hypervisor's ability to manipulate bits in the \glspl{NPT}. This way we could no longer provoke page faults, but only
    observe page faults that are ``naturally'' triggered by the \gls{VM}.
    This would however most likely need major architectural changes, like instruction set extensions.
    On the other hand,
    it would make our attack significantly harder or even infeasible, depending on the availability
    of intercepted instructions as well as the RAM size of the \gls{VM}.
    
    For the stack detection gadget, we could still use the same general strategy. But, since
    we are no longer able to provoke a page fault, allowing us to at least get the \gls{GFN} of the stack, we would
    now have to dump all of the \gls{VM}'s RAM that has ever been written to. 
    
    In order to implement a sync mechanism, that allows us to interrupt the \gls{VM} at precise points in time, we are now dependent on the availability of
    intercepted instructions.

    We thus conclude that mitigating the page fault side channel likely requires introducing major architectural changes. However, even in this case the hypervisor's control over physical memory could still be exploited to track the \gls{VM}'s memory usage.

    \subsection{Availability of Known Plaintext} \label{ssec:availability-of-known-plaintext}
    For our attack, we used the Linux kernel itself as a source of known plaintext. As customers most likely use a common Linux distribution, it can be assumed that they are running
    the kernel supplied by the respective distribution. 
    Furthermore, normal disc encryption setups do not encrypt the \texttt{/boot} partition from which the
    kernel gets loaded at boot, allowing the attacker to read the kernel binary in plaintext.
    
    If the entire boot image is encrypted, one can use the technique presented in 
    Werner et al.~\cite{werner2019severest}:
    They showed how to use \gls{IBS} 
    to reliably fingerprint specific application versions running in \gls{SEV-ES} secured \glspl{VM} based on the
    distance (measured by the \gls{GVA}) of executed return instructions of an application. 
    While they only evaluated their approach for user space applications, their result is also applicable to the Linux kernel. Another approach is building on a method presented in \cite{hetzelt2017security}: They show that the kernel location can be detected at runtime, by removing the execute permissions from all of the \gls{VM}'s memory pages, injecting an interrupt and observing the occurring page faults. Given the result of Werner et al., this could also be used to
    fingerprint the kernel based on the \glspl{GPA} of the interrupt handler functions.

    \subsection{Emulated Operations}
    Whether an instruction like \texttt{cpuid}
    or \texttt{rdtsc} is intercepted by the
    hypervisor can be configured in the control area of the
    \gls{VMCB} \cite{AMD2019architecture}. The \gls{VMCB} gets encrypted and integrity
    protected upon a \texttt{\#VMEXIT}. Furthermore, it is part of the initial attestation~\cite{sev-es}, so it cannot be manipulated by a malicious hypervisor.
    The high performance \texttt{cpuid}-based encryption oracle from Section \ref{sec:using-cpuid-as-an-encryption-oracle} can thus
    be mitigated by disabling interception of the \texttt{cpuid} instruction~\cite{AMD2019architecture}. 
    However, as
    already stated in Section~\ref{sec:interception}, emulation of instructions is an important virtualization feature, since it allows fine grained control over exposed hardware features as well as simulating a
    consistent environment during live migration.
    
    In theory, the \gls{GHCB} mechanism used under SEV-ES allows the \gls{VM} to inspect the hypervisor supplied results of an emulated operation before it continuous its operation.
    However, AMD's current \gls{SEV-ES} kernel does not implement such checks (cf. function \texttt{vmg\_cpuid}).
    For operations like \texttt{rdtsc}, the effectiveness of filtering is uncertain.
    In the recently released AMD-SNP whitepaper~\cite{sev-snp} AMD describes a mechanism,
    that allows the \gls{VM} to verify the result of the \texttt{cpuid} operation by using the \gls{SP} as a proxy. This is possible, as the hypervisor cannot interfere with the result of operations executed on the \gls{SP}.

    \subsection{Detection}
    Another important aspect, besides direct countermeasures, is attack detection. In the scenario of a malicious hypervisor spying
    on \glspl{VM}, detecting an attack could lead the guests to switch to another service or pursue legal matters.

    Since \gls{SEV} itself does not provide any integrity protection for the RAM content, this 
    must be done by the guest and in software. This is significantly complicated by the large number of possible injection points, and the fact that the injected code is only temporarily present.
    In addition, the program that inspects the guests RAM content in order to find changed code, cannot be certain that its
    own code is unchanged.
    
    Another approach is detecting abnormal behavior, like the unusual kernel base address when disabling \gls{KASLR}.
    However, detecting more transient abnormal behavior, like a manipulated random number generator, is quite difficult due to the large attack surface.

\section{Responsible Disclosure}
We have informed AMD of our findings. In our discussions they suggested that the recently released Zen 2 architecture uses an improved tweak generation, which is no longer 4-byte periodic and uses fresh randomness per boot, which may significantly complicate the described attacks. However, they did not implement an additional integrity protection, yet.
    
\section{Conclusion}
In this work we have shown that the lack of proper integrity protection can be exploited to execute arbitrary code within \gls{SEV-ES} secured \glspl{VM}. We have reverse engineered the new, \gls{XEX}-based encryption on updated AMD Epyc processors, and developed a method to control plaintext bytes by moving existing ciphertext blocks. After using this method for bootstrapping a 2-byte encryption oracle, we have shown how to place instructions to control 4 bytes and finally 16 bytes per plaintext block, yielding a 16-byte encryption oracle. In addition, we have shown how to abuse the emulated \texttt{cpuid} instruction to build a high performance encryption oracle. Compared to similar attacks, our attacks works with \gls{SEV-ES} and does not rely on any I/O operations.

We have discussed various countermeasures: A stronger tweak function and disabling instruction interception might significantly complicate our described attacks. However, we do not expect that a full mitigation is possible without implementing a proper integrity protection, which is able to detect modified ciphertext before decryption.

Proof of concept code is available at \url{https://github.com/UzL-ITS/SEVurity/}.

\ifAnon
\else
\smallskip\noindent\textbf{Acknowledgments}
This research was supported by DFG (Grant 427774779).
The authors 
would like to thank Alina Weber-Hohengrund for assisting with reverse engineering the tweak function and
Martin Radev for analyzing the page fault sequences at boot time.
Furthermore, the authors would like to thank the anonymous reviewers and especially our shepherd Yinqian Zhang for detailed comments and suggestions for improvement.

\fi

\newpage

\ifUsenix
  \bibliographystyle{plain}
  {\footnotesize
  \bibliography{main}
  }
\else
  \bibliographystyle{IEEEtranS}
  {\footnotesize
  \bibliography{IEEEabrv,main}
  }
\fi

\end{document}